\documentclass[aps,twocolumn,pre,showpacs]{revtex4}
\usepackage{amsfonts}
\usepackage{amsmath}
\usepackage{amssymb}
\usepackage{graphicx}
\begin{document}
\title{Macroion correlation effects in electrostatic screening and thermodynamics of highly charged colloids}
\author{R. Casta\~{n}eda-Priego$^{1}$}\email{ramoncp@fisica.ugto.mx}
\author{L. F. Rojas-Ochoa$^{1,2}$}
\author{V. Lobaskin$^{3}$}
\author{J. C. Mixteco-S\'{a}nchez$^{1}$}
\address{$^{1}$Instituto de F\'{\i}sica, Universidad de Guanajuato, Loma del Bosque 103, 37150 Le\'{o}n, Mexico}
\address{$^{2}$Departamento de F\'{\i}sica, Cinvestav, Av. Instituto Polit\'{e}cnico Nacional 2508, 07360 M\'{e}xico D. F., Mexico}
\address{$^{3}$Physik-Department, Technische Universit\"at M\"unchen, James-Franck-Str., D-85747 Garching, Germany}

\begin{abstract}
We study macroion correlation effects on the thermodynamics of highly charged colloidal suspensions using a mean-field theory and primitive model computer simulations. We suggest a simple way to include the macroion correlations into the mean-field theory as an extension of the renormalized jellium model of Trizac and Levin [Phys. Rev. E {\bf 69}, 031403 (2004)]. The effective screening parameters extracted from our mean-field approach are then used in a one-component model with macroions interacting via Yukawa-like potential to predict macroion distributions. We find that inclusion of macroion correlations leads to a weaker screening and hence smaller effective macroion charge and lower osmotic pressure of the colloidal dispersion as compared to other mean-field models. This result is supported by a comparison to primitive model simulations and experiments for charged macroions in the low-salt regime, where the macroion correlations are expected to be significant.
\end{abstract}
\maketitle

\section{Introduction}
Structure and thermodynamics of charged colloidal dispersions
became a subject of a renewed interest over the last decades due
to development of experimental and theoretical techniques and
accumulation of new data incompatible with classical theories
\cite{hansen,89,netz}. Considerable theoretical efforts have been
invested into an upgrade of existing mean-field approaches with an
inclusion of additional correlation effects such as counterion or
macroions correlations, which are missing in the classical
Poisson-Boltzmann theory.

Whereas in the aqueous dispersions of micrometer-sized particles
the correlations of monovalent counterions are usually negligible,
it is known that they might become important for small strongly
charged macroions \cite{vlachy,lobaskin:99}. The role of
counterion correlations has been extensively studied by various
means, including integral equation theories and molecular
simulations starting from eighties
\cite{Perbo,14,Ample,Ample2,vlachy,kjellander1,kjellander2} and is
currently well understood
\cite{vlachy,kjellander1,kjellander2,89,netz}. In contrast,
macroion correlation effects are usually not included in the
mean-field approaches and therefore not quantified. One can expect
these effects to be significant in systems with thick double
layers, such as deionized colloidal dispersions.

In order to specify our interest in macroion correlation effects
we would like to start from a simple energy argument. Charged
colloidal suspensions are composed of a large number of particles
of different types. If the molecular details of the solvent and
dielectric discontinuities are neglected, one arrives to the
primitive electrolyte model. On this level, a charged colloidal
dispersion is an asymmetric electrolyte consisting of strongly
charged macroions and small counterions. In addition, at least two
different species of salt ions are usually present. A
straightforward application of the Debye-H\"uckel-like mean-field
description is usually not successful due to strong spatial
correlations of different ionic species. To deal with the
correlations, one can attempt to construct an hierarchy of
interactions from a quite general viewpoint. If we look at a
system of macroions and small ions (including counterions and salt
ions) we can divide the contributions to the potential energy into
three categories: Macroion-macroion (MM), macroion-ion (MI), and
ion-ion (II). The relative importance of these terms can be
estimated based on simultaneous consideration of their magnitudes
and distances, on which they are operating. Macroions repel each
other, so that their interaction is of the order of $\beta
u_{MM}\approx \lambda_B Z_M^2 \exp{(-\kappa d)}/d$, the
macroion-ion contribution $\beta u_{MI} \approx \lambda_B Z_M Z_I
\exp{(-\kappa a)}/a$, and the ionic part $\beta u_{II} \approx
\lambda_B Z_I^2\exp{(-\kappa d_{I})}/d_{I}$. Here, $\lambda_B =
e^2/(4 \pi\varepsilon \varepsilon_0 k_B T)$ is the Bjerrum length,
$\beta^{-1}=k_B T$ the inverse of the thermal energy, $k_B$ the
Boltzmann constant, $T$ the temperature, $Z_M, Z_I$ macroion and
ion valence respectively, $a$ the macroion radius, $\kappa = 4
\pi\lambda_B (Z_I^2 c_{s} + n Z_M)$ the screening parameter,
$d=n^{-1/3}$ and $d_{I}=c_{s}^{-1/3}$ the mean macroion-macroion
and ion-ion distance respectively, $c_{s}$ the salt content and
$n$ the macroion number density. Setting $Z_M = 1000, Z_I= \pm 1$,
$c_{salt}= 1$ mM, $a = 100$ nm and the macroion volume fraction to
$0.01$, we get $\beta u_{MM} \approx 3 \times 10^{-30}$, $\beta
u_{MI} \approx3 \times 10^{-4}$, and $\beta u_{II} \approx 3
\times 10^{-2}$. From this naive estimate, one can conclude that
at $\kappa a \gg 1$ the two last contributions dominate the system
thermodynamics. When the charge sign is taken into account, the
negative MI contribution starts prevailing in the total energy, as
the significant part of II contribution; consisting of nearly
equal number of terms of opposite signs, cancels itself out. Due
to strong screening, the thermodynamic properties of such
dispersion do not differ much from a simple electrolyte, except
for small layer of thickness $\kappa^{-1}$ around the macroion
surface. One observes a much different picture in the regime of
thick double layers $\kappa a \ll 1$. A similar estimate for
$c_{salt}= 1 \mu$M gives $\beta u_{MM} \approx 80$, $\beta u_{MI}
\approx 5$ and $\beta u_{II} \approx 4 \times 10^{-3}$. If this
energy per ion is weighted by the corresponding number of species,
the MI contribution dominates so that the total Coulomb energy
becomes negative \cite{lobaskin:99}. We can also see that the II
interactions are unlikely to influence the dispersion properties
in case of monovalent ions. The first two contributions, however,
have to be taken into account. The common method of dealing with
this situation involves: (i) solving the Poisson-Boltzmann (PB)
equation for a single macroion, (ii) renormalization of the MM
interaction parameters based on the Debye-H\"uckel-like
approximation for the long distance part of the double layer, and
(iii) solution of the one-component MM model with an effective
interaction potential $u_{\text{eff}}$
\cite{Alexander,Belloni,lobaskin:99}. This procedure respects the
leading role of the MI interaction, while the effect of MM
correlations in this approach enters only on the level of the
one-component (OCM) description. The latter approach can be
modified in different ways to account for more pronounced role of
macroion correlations and the resulting charge inhomogeneities
using the Wigner-Seitz cell model \cite{Belloni}. In this case,
the structure of the double layer reflects the inhomogeneous
macroion (and hence the counterion) distribution via the cell
construction. Numerical schemes based on the cell model and charge
renormalization have been successful in describing properties of
charged colloidal dispersions
\cite{Alexander,Belloni,lobaskin:99}.

An alternative route to include macroionic contribution into
electrostatic screening is based on the jellium approximation for
macroions \cite{Jellium,Jellium1}. Although the range of validity
of the jellium model might be limited to weakly correlated
macroion systems, this model can be easily extended to different
situations, i.e. rod-like colloids \cite{salete06} or asymmetric
electrolytes and, particularly, its equation of state takes a
simple analytical form (see equation (\ref{eq2b}) below).

In this work, we follow the jellium description, which we improve
using a simple construction that introduces spatial macroion
correlations. The main goal of the present work is therefore to
study the effect of macroion correlations on the parameters of the
OCM and the corresponding dispersion thermodynamics. Another issue
we would like to address is the calculation of the equation of
state of the dispersion. It is well known that recovering the
accurate equation of state basing on the OCM representation is
problematic, as the ionic degrees of freedom are omitted from the
description and wall effects are not included on the macroion
virial contribution \cite{Dobnikar05}. At the same time, the
thermodynamic properties can be easily extracted from the same
two-component (or multi-component) description that is used for
calculation of the OCM effective charge and screening length, i.e.
directly from the WS cell or jellium model
\cite{Alexander,Belloni,LobaskinJCP,Jellium,Jellium1,Stevens,Trizac}.
These models are usually solved using the non-linear PB equation
including only one colloidal particle, or MC simulation
\cite{Alexander,lobaskin:99,lobaskin:01,wenner:82}. In this work
we will use non-linear PB equation with the jellium boundary
conditions.

\section{Macroion correlations on the mean-field level}

Although the PB cell and the renormalized jellium models do not
address the macroion degrees of freedom, they implicitly include a
model of macroion distribution. The cell model supposes well
separated particles, where the role of the neighboring macroions
consists of limiting the volume available for small ions while
$g_{MM}(r)$ is simply zero inside this cell. The double layer
inside the cell is otherwise unperturbed by the rest of the
system. In contrast, the jellium model assumes $g_{MM}(r) = 1$ for
$r>2a$(diameter of the sphere), i.e. an ideal gas of macroions. As
we already noted in the introduction, this approximation might be
unsatisfactory for low-salt colloidal dispersions, where macroions
strongly repel each other even at the mean interparticle
separation. We therefore will try to avoid considering uniform
macroion distributions. On the simplest level, the uniform
distribution can be replaced by a $g_{MM}(r)$ taken in the form of
a step function. This choice is motivated by the observation that
a charge-stabilized colloidal suspension at low salt shows a
highly structured $g_{MM}(r)$ with a characteristic lengthscale
described by the mean interparticle distance $d=n^{-1/3}$
\cite{Klein}. In particular for $r<d$, $g_{MM}(r)$ is almost zero,
a feature that is known as the "correlation hole". The total
charge density in the system at a distance $r$ from the center of
a macroion becomes
\begin{equation}
\label{rho}\rho(r) = -Z_{\text{eff}} e  n g_{MM}(r) + e\rho_+(r) - e\rho_-(r),
\end{equation}
where $g_{MM}(r)= 0$ for $r < d$, and $g_{MM}(r)=1$ for $r \geq
d$; $\rho_\pm(r)$ are the concentrations of salt cations and
anions and $e$ is the elementary charge. We should stress that the
"correlation hole" approximation for $g_{MM}(r)$ can be justified
only for low-salt systems where the main peak position of
$g_{MM}(r)$ scales with $n^{-1/3}$. In a more general case, one
should consider a "correlation hole" of size $d^{*}$ such that the
main peak position scales with it and which should be valid for
higher salt concentrations or weakly charged macroions. This point
will be addressed elsewhere \cite{Mix06}.
\begin{figure}
\begin{centering}
\includegraphics[width=0.45 \textwidth]{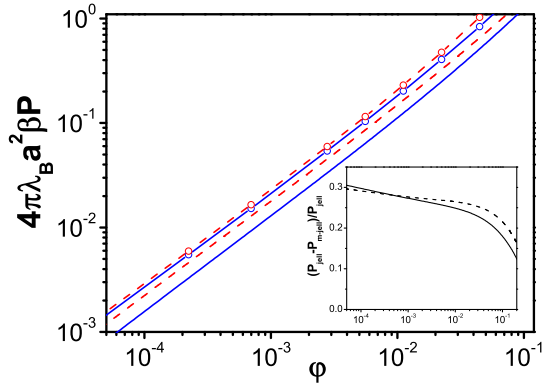}
\caption{\label{fig0} Effect of the macroion correlation
correction to the renormalized jellium model on the reduced
pressure in a salt-free dispersion. The solid and dashed curves show the
pressures in the moderate effective charge regime ($Z_M
\lambda_B/a = 10$) predicted by the jellium and m-jellium, respectively and the solid and dashed curves with symbols the saturated effective charge regime ($Z_M\lambda_B/a = 1000$). The inset shows the
relative difference between the renormalized jellium and m-jellium
pressures. This relative difference describes also the change in
the effective charge in the salt-free system.}
\end{centering}
\end{figure}

Our model includes now the major part of macroion correlations by
placing a macroion at the center of its correlation hole with a
size that depends explicitly on the concentration. The macroion
distribution outside this hole is still approximated by an ideal
gas of macroions, as in the jellium approach \cite{Jellium}. We
enforce the smeared-out background charge, $Z_{\text{back}}$,
representing the other macroions around the tagged macroion to
coincide with the effective charge, $Z_{\text{eff}}$, as in the
original work of Trizac and Levin \cite{Jellium1}. The
$Z_{\text{back}}$ is determined by the electroneutrality condition
for the total charge density in the bulk, $2 c_s \sinh
\left[e\phi(\infty) /k_B T \right] = n Z_{\text{back}}$, where
$\phi(\infty)$ represents the electrostatic potential in the bulk.
The electroneutrality condition also allows us to determine the
effective screening parameter, $\kappa^{2}_{\text{eff}}= 4
\pi\lambda_B \sqrt{Z_{\text{eff}}^{2} n^{2} + 4 c_{s}^{2}} $
\cite{Jellium1}. We therefore call our model a modified jellium
model (m-jellium).

\begin{figure}
\begin{centering}%
\mbox{}\\[1cm]
\includegraphics[width=0.45 \textwidth]{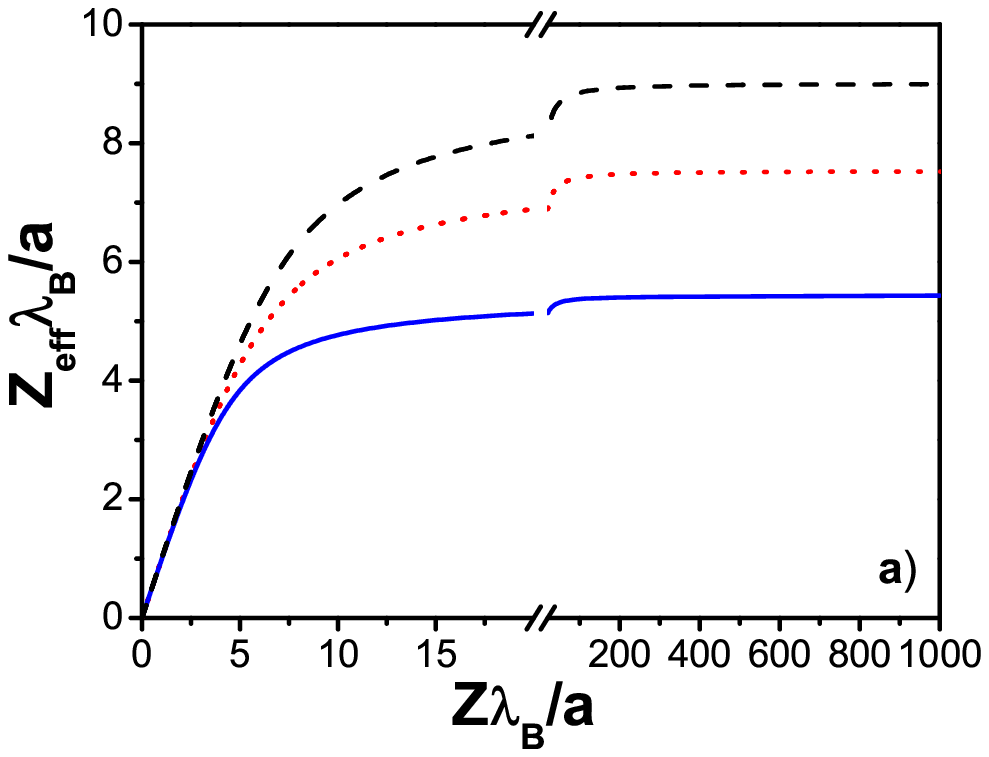}
\includegraphics[width=0.45 \textwidth]{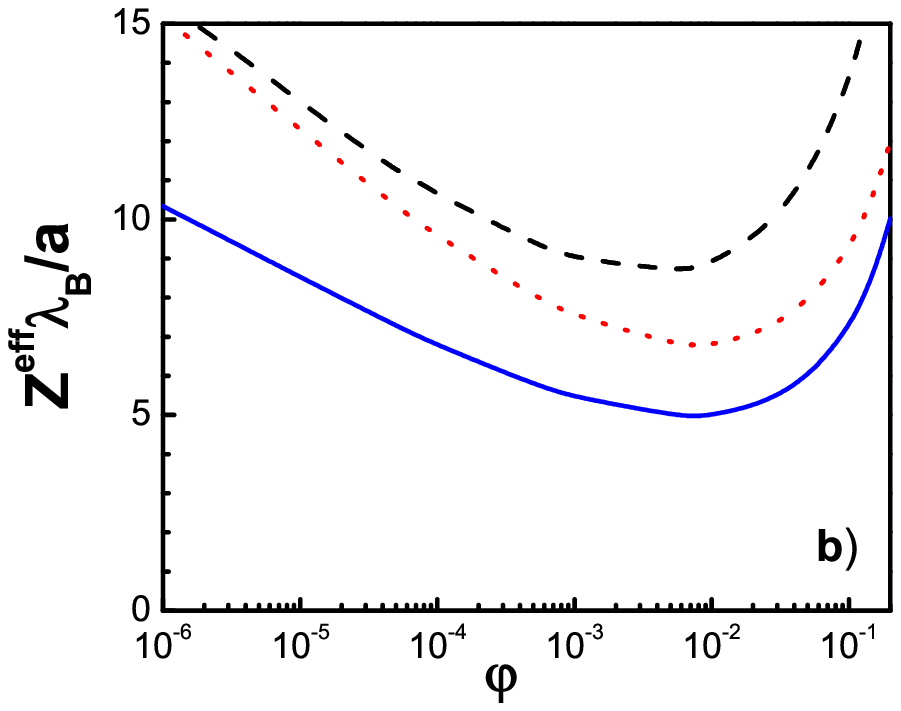}
\caption{\label{fig1}a) Effective charge for a salt-free suspension at a volume fraction of $\varphi=0.001$ as a function of the bare charge and b) effective charges at saturation for charge-stabilized colloidal suspensions, salt-free case, as a function of volume fraction. In both cases, dashed lines are for the PB-cell model, dotted lines for the jellium model and solid lines for the modified jellium approach.}
\end{centering}
\end{figure}

The effect of the introduced modification is illustrated in Fig.
\ref{fig0}. One can see that the pressure in a salt-free colloidal
dispersion becomes smaller in the m-jellium model as compared to
the original jellium result at all macroion volume fractions. For
both highly ($Z_M\lambda_B/a = 1000$) and moderately charged
systems ($Z_M \lambda_B/a = 10$), the relative difference between
the two models is maximal (about 0.3) at the lowest volume
fraction and decreases as the volume fraction increases (it does
not exceed 0.15 at $\varphi=0.1$). The relatives differences
between pressures from both models are illustrated in the inset of
Fig. \ref{fig0}. The pressure in the salt-free case is
proportional to the effective charge, equation (\ref{eq2b}), so
the effective charge variation in both models is also described by
the same equation. The difference between both models decreases
with density because the size of the correlation hole also scales
as $n^{-1/3}$. Therefore, it is expected that results from both
models will coincide at higher volume fractions.

For comparing different systems it is convenient to present the
effective charge in the form $Z_{\text{eff}}\lambda_{B}/a$. In
Fig. (\ref{fig1}a) we compare the effective macroion charge as a
function of its bare charge for the PB-cell, jellium and m-jellium
models in the salt-free case. Each model predicts different values
of the effective charges at saturation ($Z_{\text{bare}} \to
\infty$). However, for small bare charges all of them recover the
correct limiting behavior $Z_{\text{eff}}=Z_{\text{bare}}$. It is
interesting to note that for the salt-free case at saturation the
system properties are determined by only one parameter: the
macroion volume fraction, $\varphi=4\pi a^{3} n/3$. In Fig.
(\ref{fig1}b) we compare the saturated effective charges obtained
from each model. The behavior of $Z_{\text{eff}}$ in the range
$\varphi \lesssim 10^{-2}$ can be understood by a compression of
the ionic double layers. As $\varphi$ increases, counterions are
pushed towards the macroion surfaces thus reflecting the win of
the entropy over the energy and therefore leading to a gradual
decrease of $Z_{\text{eff}}$ until a well-defined minimum appears
around $\varphi \thickapprox 10^{-2}$ and we observe the growth in
the effective charge at $\varphi>10^{-2}$. This reduction in the
effective charge in the m-jellium as compared to the original
jellium model follows from the weaker screening at $r<n^{-1/3}$
(now the macroions are excluded from the internal parts of the
double layer), which leads to a higher free energy cost of
charging the macroion. In other words, the weaker screening result
to stronger attraction of the counterions to the macroion surface.
\begin{figure}
\begin{centering}
\includegraphics[width=0.45 \textwidth]{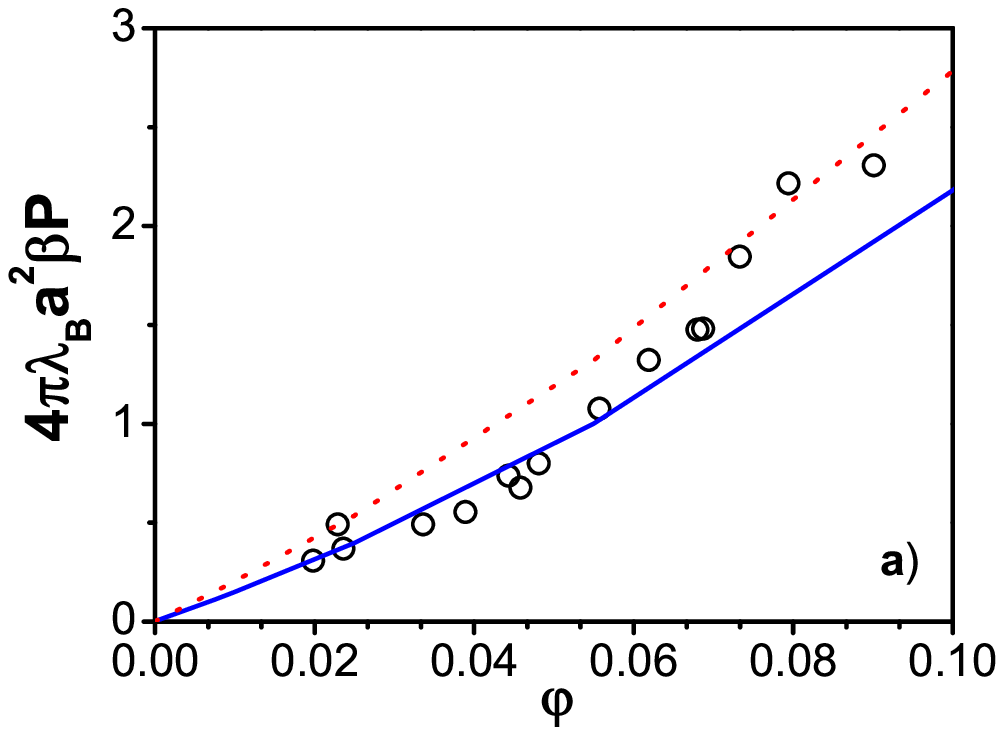}
\hfill \includegraphics[width=0.45 \textwidth]{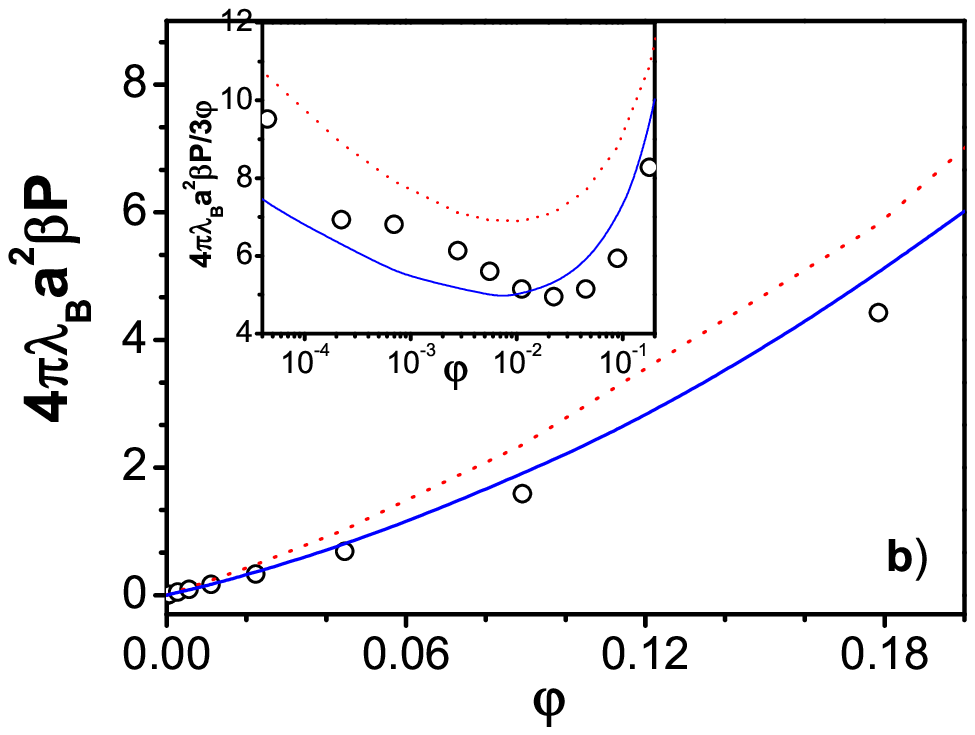}
\caption{\label{fig2} a) Pressure as a function of the volume
fraction obtained from the m-jellium approach (solid line) and the
renormalized jellium approximation (dashed line). Symbols
correspond to the Reus' experiments \cite{Reus}. No adjustable
parameters have been used and effective charges at saturation are
considered. b) Pressure as a function of the volume fraction
obtained from the m-jellium approach (solid line) and from the
renormalized jellium approximation (dashed line) with
$Z_{\text{bare}\,}\lambda_B/a=19.47$ and $\lambda_B/a=0.3245$.
Symbols correspond to primitive model simulations.}
\end{centering}
\end{figure}

\section{Pressure}
\label{Pressure}

Within the jellium model, the equation of state reads \cite{Jellium1}
\begin{equation}
\label{eq2b}
\beta P = n  + \sqrt{Z_{\text{eff}}^{2}n^{2}+4c_{s}^{2}}.
\end{equation}
After exclusion of the condensed counterions by the charge
renormalization procedure only the free ions contribute to the
pressure. In the salt-free case, $c_{s}=0$, the equation of state
given by Eq. (\ref{eq2b}) simply reduces to $\beta
P=n(1+Z_{\text{eff}})$. In the low electrostatic coupling regime
(where $Z_{\text{eff}}$ coincides with $Z_{bare}$), equation
(\ref{eq2b}) recovers (see also Fig. \ref{fig1}) the ideal gas
pressure $\beta P \simeq \rho_c (1+Z_{bare})$ \cite{Belloni}. In
Fig. (\ref{fig2}a) we compare predictions from the renormalized
and the m-jellium models for the osmotic pressure data from Reus
etal. \cite{Reus}, obtained for deionized suspensions of
bromopolystyrene particles. Although the overall agreement between
experiments and both models is good, for $\varphi<0.07$ the
m-jellium approach performs visibly better while for higher
densities the renormalized jellium model seems to be a better
approximation. Further on, in Fig. (\ref{fig2}b) we compare the
results from both jellium-like models with data from primitive
model simulations for salt-free asymmetric electrolyte with an
asymmetry in charge $60:1$ (we use the same cluster MC simulation
protocol and settings as in Ref. \cite{lobaskin:99} with 80
macroions). Here we observe that both models reproduce the
pressure behaviour qualitatively, while the m-jellium gives a
better quantitative agreement in the wide range of macroion
concentrations. An alternative representation of the same data is
shown in the inset of Fig. (\ref{fig2}b), where both models are
compared to the osmotic coefficients for the same simulated
system. From this representation, it is clear that the m-jellium
describes the results of numerical simulations better. Thus, our
model improves the osmotic pressure for suspensions in the
counterion-dominated screening regime.

\section{Structure}

The effective charge and the screening parameter computed from our
m-jellium can be used to calculate the effective pair potential
and the structure of the suspension. We assume the effective pair
interaction between macroions to have the Yukawa form \cite{Belloni}
\begin{equation}
\label{eq1}
\beta u_{\text{eff}}(r) = Z_{\text{eff}}^{2} \lambda_B \left[\frac{\exp(\kappa_{\text{eff}}\, a)}{1 + \kappa_{\text{eff}}\, a}\right]^2 \frac{\exp(-\kappa_{\text{eff}}\, r)}{r}.
\end{equation}
The pair distribution of the macroions interacting through the
effective pair potential $(\ref{eq1})$ can be computed using the
Ornstein-Zernike (OZ) equation \cite{oz},
\begin{equation}
\label{oz}
h_{MM}(r)=c^{\text{eff}}(r)+n\int d^{3}r^{\prime}c^{\text{eff}}(r^{\prime}) h_{MM}(\left\vert\mathbf{r}-\mathbf{r}^{\prime}\right\vert),
\end{equation}
where $h_{MM}(r)=g_{MM}(r)-1$ and $c^{\text{eff}}(r)$ is the
so-called effective direct correlation function
\cite{Belloni,rcp:jpcm03}. An additional closure relation is
needed to solve the OZ equation. In particular, the Rogers-Young
(RY) closure relation \cite{RY} can be used to solve the OZ
equation self-consistently. The RY closure enforces both
thermodynamic and density fluctuations to be the same in order to
calculate both the structure and thermodynamics of a colloidal
suspension and it is known to accurately describe Yukawa systems
\cite{Klein}. Technically, this is done by computing the
isothermal compressibility using the virial route for the OCM,
$\chi_{v}^{-1}=(\partial \beta P_{\text{MM}}/\partial n)_{T}$,
where $P_{\text{MM}}$ is the macroion--macroion virial
contribution and can be written as
\begin{equation}
\label{eq:ocmvirial}
\beta P_{\text{MM}}=\rho_{c}-\frac{\rho_c^2}{6} \,\int_{r=2a}^\infty g(r)\, \frac{d\beta u_{\text{eff}}(r)}{dr}\,r\,d^3\vec{r}.
\end{equation}
The isothermal compressibility can be also computed through the
relation $\chi_{c}^{-1}=1-n \tilde{c}^{\text{eff}}(q=0)$
\cite{Belloni}, where $\tilde{c}^{\text{eff}}(q)$ is the Fourier
transform of the effective direct correlation function. Then, the
RY closure relation enforces both routes to give the same
isothermal compressibility $(\chi_{v}=\chi_{c})$ in order to
guarantee, at least partially, the thermodynamic consistency
\cite{RY}.

It is important to note that the OCM pressure is usually very
different from the pressure measured in the full multicomponent
electrolyte. This discrepancy follows from the dominance of
microion contribution to the pressure. In fact, the total pressure
can be well approximated by the partial pressure of the small ions
measured using the contact value theorem at the WS cell boundary
or at infinity \cite{lobaskin:99}. Moreover, as we have seen in
the last section, equation (\ref{eq2b}) provides an excellent
equation of state for highly charged colloids. It is now clear
that equation (\ref{eq:ocmvirial}) is not able to reproduce the
isothermal compressibility of the suspension and thus it cannot be
used to reproduce the structure of the suspension. Nevertheless,
if the total pressure (or compressibility) of the system is known
from the original system (or a solution of the multicomponent
m-jellium model), then it can be combined with the OCM to get the
thermodynamically consistent description. Thus, the OCM pressure
should be replaced with the pressure in the full system, which is
given by Eq. (\ref{eq2b}) within our mean-field approach. We
stress that the pressure in our procedure is evaluated from the
same mean-field parameters that give the correct macroion
structure. This remark concerns also the osmotic compressibility
$\chi_{T}$, which is related to the infinite wavelength limit of
the macroion structure factor \cite{Belloni}. Knowing the macroion
structure is therefore enough to compute the correct osmotic
compressibility of the multicomponent system. On the other hand,
the isothermal osmotic compressibility can be calculated with
$\chi_{v}^{-1}=(\partial\beta P/\partial n)_{T}$, where $P$ is
given by Eq. (\ref{eq2b}), with the same result.

We would like to note also that these simple ideas can be applied
for interpretation of experimental data. The effective macroion
charge is usually extracted from the structure factors measured in
scattering experiments through fitting the curve with a OZ-RY
scheme with an effective Yukawa potential. A reverse version of
our approach can be used to predict accurately the equation of
state of colloidal suspensions at low salt via the relation
between the effective charge and the osmotic pressure.
\begin{figure}
\begin{center}
\includegraphics[width=0.45 \textwidth]{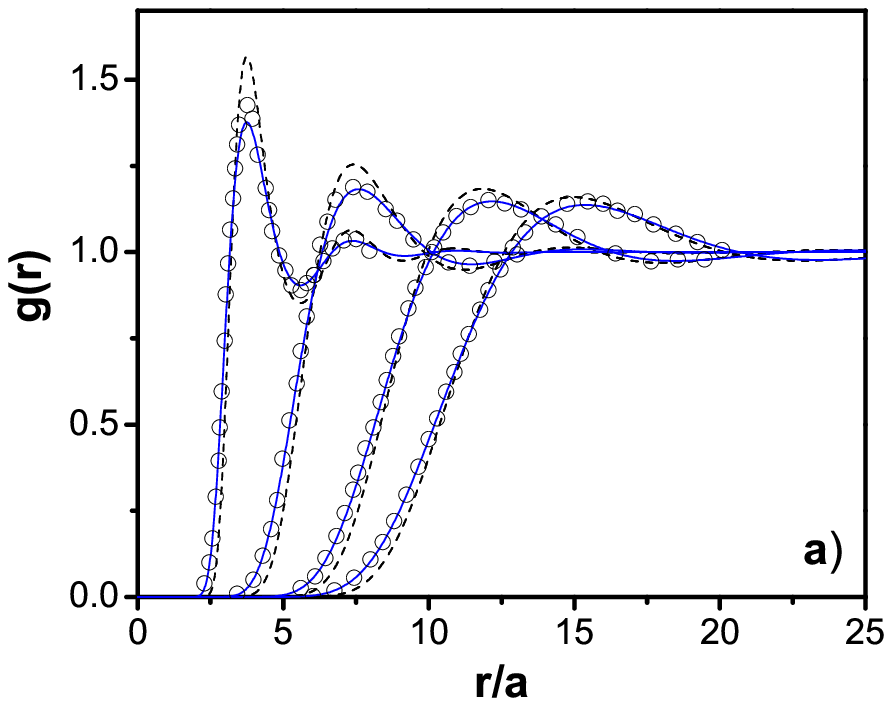}
\includegraphics[width=0.45 \textwidth]{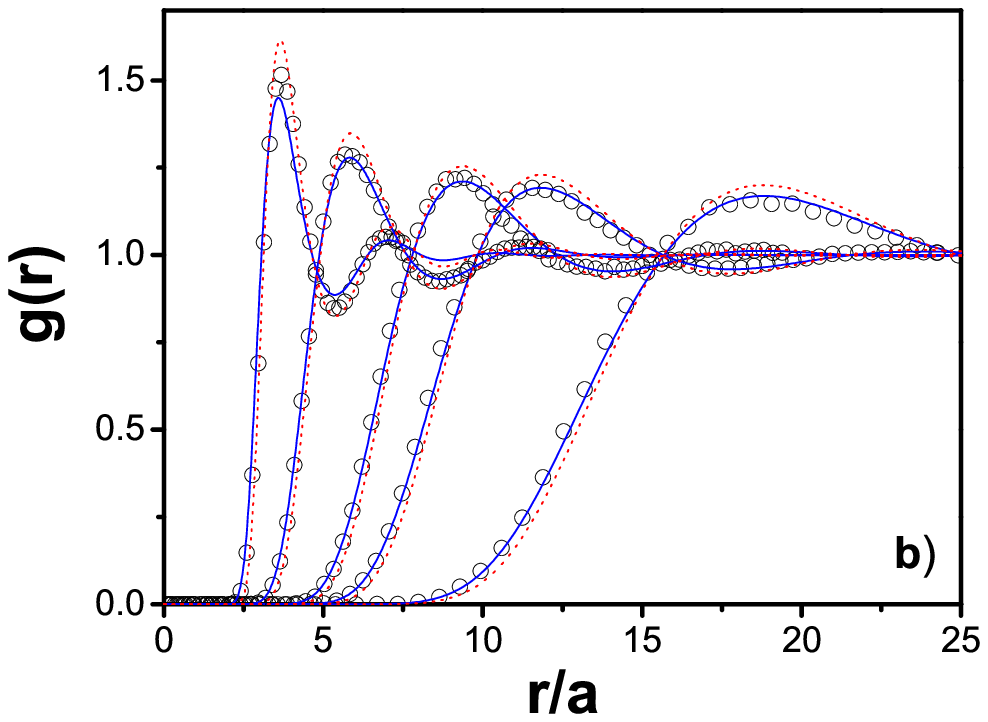}
\caption{\label{fig3} Macroion-macroion pair-correlation functions
from primitive model simulation (symbols) and from our OZ-RY
scheme (see text) with effective m-jellium parameters (solid
lines). a) From the standard OZ-RY scheme (see text) with the OCM
equation of state and PB-cell parameters (dashed lines), and b)
from our OZ-RY scheme with effective jellium parameters (dotted
lines). From left to right, the packing fractions are a)
$\varphi=0.08$, $0.01$, $0.0025$, $0.00125$, and b)
$\varphi=0.089$, $0.022$, $0.0055$, $0.0027$, $0.00069$.}
\end{center}
\end{figure}

In figure (\ref{fig3}a) we compare the radial distribution
functions (RDF) from our primitive model simulations (symbols)
with numerical results obtained using our OZ-RY scheme with the
m-jellium parameters (solid lines). Also, results from the
standard RY route \cite{RY} and screening parameters from the
PB-cell model are provided (dashed lines). For the sake of
clarity, the results with parameters from the renormalized jellium
through the standard route are not shown, however it has found
that they underestimate the structure (for a more detailed
analysis see Ref. \cite{Dobnikar06}). The highest macroion charge
taken in simulations ($Z_{\text{bare}}=60$) does not bring the
system into fully saturated effective charge regime. The value for
$\lambda_B/a$ considered (among others) was $0.324$. In figure
(\ref{fig3}a) we clearly observe a good agreement between
simulations and m-jellium results although still small differences
around the main peak of the RDF for the higher volume fractions
($\varphi>0.04$) can be observed. These small differences might
result from the macroion shielding effect, which usually rises the
macroion distribution peak in dense suspensions \cite{lobaskin03}.
However, our results show a visibly better agreement with
simulations than those obtained with other mean-field approaches
(dashed lines). In figure (\ref{fig3}b) we compare the RDF from
the primitive model simulations described in section
\ref{Pressure} with numerical results obtained using our OZ-RY
scheme with both m-jellium and jellium parameters. We observe that
the jellium model always overestimates the structure of the
suspension while m-jellium shows a better agreement. It is
remarkable that the behavior of $g_{MM}(r)$ in figure (\ref{fig3})
is accurately reproduced by our scheme in the whole range of
distances, whereas other mean-field schemes overestimate the
short-range behaviour of $g_{MM}(r)$. Also, we note that at high
densities the mean peak of the pair correlation is predicted less
accurately. 

\section{Conclusions}

We have introduced a procedure of including macroion correlations
into a mean-field theory of screening in charged colloidal
dispersions, which leads to modification of the effective
parameters of the OCM: the macroion effective charge and the
screening length.

The evaluation of colloidal effective charges in suspensions with
weak screening is based on the solution to the nonlinear PB
equation for the electric double layer in the presence of other
macroions. Our model represents a modification of the renormalized
jellium approach by Trizac and Levin. The obtained results suggest
that these correlations are important in systems with thick double
layers such as deionized colloidal suspensions, i.e. in the
counterion-dominated screening regime.

Our model describes well experimental results of thermodynamic
quantities in colloidal dispersions, such as the osmotic pressure.
The static structure of colloidal suspensions is also accurately
reproduced by an OCM scheme that uses the screening parameters of
the m-jellium model. Moreover, from the experimental point of
view, the effective parameters of the RY integral equation scheme
can be used to predict accurately the equation of state via the
relation between the effective charge and the osmotic pressure in
our model.

\begin{acknowledgments}
It is a pleasure to thank to E. Trizac and H. H. von Gr\"{u}nberg
for fruitful discussions and valuable comments on the manuscript.
We also thank to PROMEP, CONACyT-Mexico (grants 46373/A-1 and
51669) and CONCyTEG for financial support.
\end{acknowledgments}

\end{document}